\documentclass[twocolumn,preprintnumbers,amsmath,amssymb,amsfonts]{revtex4}

\usepackage{graphicx}
\usepackage{mathrsfs}
\usepackage{color}
\usepackage{mathtools}
\usepackage{float}

\newcommand{\bra}[1]{\left\langle{#1}\right|}
\newcommand{\ket}[1]{\left|{#1}\right\rangle}

\begin{document}


\title{Complexity of quantum states in the stabilizer formalism}
\author{Shuangshuang Fu}
\affiliation{%
	School of Mathematics and Physics, University of Science and Technology Beijing, Beijing 100083, China
}%

\author{Shunlong Luo, Yue Zhang}
\email{zhangyue115@amss.ac.cn}
\affiliation{State Key Laboratory of Mathematical Sciences, Academy of Mathematics and
	Systems Science, Chinese Academy of Sciences,  Beijing 100190,
	China\\
	School of Mathematical Sciences, University of Chinese
	Academy of Sciences, Beijing 100049, China}
\begin{abstract}

We initiate an investigation into a notion of state complexity for discrete-variable quantum systems. Specifically, we propose an information-theoretic quantifier for the complexity of quantum states within the stabilizer formalism of quantum computation. This is achieved by leveraging the symmetric Jordan product (associated with classicality) and the skew-symmetric Lie product (linked to quantumness) between the square root of the quantum state and the Heisenberg-Weyl displacement operators. We establish the fundamental properties of this quantifier and demonstrate that state complexity is closely related to the nonstabilizerness of quantum states via the  $L^4$-norm  of their characteristic functions.


\end{abstract}

\maketitle
\setlength{\parindent}{1em}

\section{Introduction}
The notion of complexity encompasses diverse connotations across scientific disciplines, ranging from computational cost and statistical complexity to structural intricacy and beyond \cite{Cook1983,Li2008,Shen2017,Bu2023}. In our approach to complexity, for a classical system described by a probability distribution function, the statistical measure of complexity should embody the interplay between the information stored in the system and its disequilibrium. A class of significant quantities of complexity are defined as products of two scalar quantities, which capture the two complementary features of the probability distribution function \cite{Comp1995,Compexity2012}. Some information-theoretic quantities, including the various entropies and Fisher information, are employed in the quantification of complexity. 

In quantum theory, the concept of complexity has been explored from a multitude of perspectives \cite{Bras2003,Mont2012,Cubi2012,Compexity2012,Leve2018,Arai2023,Nura2025,Li2025com,Tang2025,Tang20251,T3,Varg2026}. Recently, complexity measures tailored to continuous-variable quantum systems have been put forward \cite{Tang2025,Tang20251,T3}. The principal approach to formulating a complexity measure typically entails two key steps: first, defining a complexity measure for a probability density function; second, associating a probability density function with the target quantum system. In the realm of quantum information, complexity seeks to characterize the delicate trade-off between classical and nonclassical properties. For example, for continuous-variable quantum systems, Tang et al. \cite{Tang2025,Tang20251} leveraged the Husimi distribution to propose a complexity measure based on the Wehrl entropy and the Fisher information. The Wehrl entropy quantifies the spread of the distribution, whereas the Fisher information encodes the complementary behavior. However, the complexity of discrete-variable quantum systems has been far less extensively investigated \cite{Varg2026}.

The stabilizer formalism for discrete-variable quantum systems constitutes a powerful framework to investigate quantum error correction and universal fault-tolerant quantum computation \cite{Niel2010}. Within this formalism, quantum states are categorized into stabilizer states and magic states (non-stabilizer states) \cite{Niel2010,Gott1997,Gott1998}. Stabilizer states, which act as ``classical objects'', alongside Clifford operations, can be efficiently simulated on classical computers, a result formalized by the Gottesman-Knill theorem \cite{Gott1997,Gott1998}. In contrast, magic states are indispensable to realize universal quantum computation and to exhibit a quantum computational advantage \cite{MGQuantifier2014,Brav2016,Howa2017,Ahma2018,Hein2019,Wanf2019,Liu2022,Leon2022,Labi2022,Dai2022,Feng2022,Haug2023,Haug20232,Li2025,Zhan2026}.


Parallel to this in quantum measurement theory, the so-called symmetric informationally complete positive operator-valued measures (SIC-POVMs) are an important class of quantum measurements \cite{Rene2004,Appl2005,Zhu2010,Zaun2011,Samu2025}, whose existence in all dimensions remains a famous open problem (Zauner’s conjecture) \cite{Zaun2011}. The SIC-POVMs play a crucial role in both theoretical studies and applications of quantum measurements \cite{SIC2006,SIC2010,SIC2011,SIC2015,SIC2018,SIC2019,SIC2020}. The fiducial states, from which SIC-POVMs are generated \cite{Zaun2011}, are known to possess the maximal nonstabilizerness (magic) \cite{Dai2022,Feng2022} and be far from stabilizer states \cite{Appl2014,Ande2015,Fu2025}, promising interesting applications in quantum computation.

In this work, we  address a resource-theoretic manifestation of complexity within the stabilizer formalism: we aim to quantify complexity through the interplay of commutation and anti-commutation of operators. More specifically, we introduce an  information-theoretic quantifier of complexity constructed from complementary quantities derived from the Jordan product (anti-commutator, capturing ``classical" spread) and Lie product (commutator, capturing ``quantum" disturbance) of the square root of a quantum state with the Heisenberg-Weyl (displacement) operators. The Lie product term in the proposed expression can be regarded as a generalized Wigner-Yanase skew information—a type of quantum Fisher information—thus capturing the quantum feature. In contrast, the Jordan product term captures classical features.

This work is organized as follows. In Sec. \ref{s2}, we review some key preliminary concepts, including the stabilizer formalism, the characteristic functions, the SIC-POVMs, and the fiducial states. In Sec. \ref{s3}, we first establish the trade-off relation between the Jordan product and Lie product of the square root of a quantum state with Heisenberg-Weyl displacement operators, then we propose a complexity quantifier within the stabilizer formalism, investigate its basic properties, and elaborate on its complementary relation to magic. We illustrate the quantifier of complexity in the qubit system in Sec. \ref{s4}. Finally, we draw our conclusions in Sec. \ref{s5}. 

\section{Preliminary}\label{s2}
	
\subsection{Stabilizer formalism}

For a $d$-dimensional quantum system $\mathbb{C}^d$ with computational basis $\{|j\rangle : j\in \mathbb{Z}_d\},$ where $\mathbb{Z}_d=\{0,1,\dots, d-1\} $ is the ring of integers modulo $d,$ the shift operator $X$ and the phase operator $Z$ are defined as \cite{Gott1998,Gott1997}
\begin{eqnarray*}
	X|j\rangle = |j+1 \rangle , \qquad   Z|j\rangle = \omega ^{j} |j\rangle , \qquad \omega={e}^{2\pi i/d},\ j\in\mathbb Z_d.
	\label{XZ}
\end{eqnarray*}
Employing the shift and phase operators as the building blocks, the discrete Heisenberg-Weyl operators are defined as \cite{Appl2005}
\begin{equation}
	D_{k,l}=\tau^{kl}X^k Z^l, \qquad \tau=-{e}^{\pi i/d}, \ k,l\in \mathbb{Z}_d.  \label{HW}
\end{equation}
Obviously, the Heisenberg-Weyl operators are unitary operators and $D_{0,0}={\bf 1}$ is the identity operator.  Moreover, they obey the following multiplication relations
\begin{equation*}\label{hwsr}
	D_{k,l}D_{s,t}=\tau^{ls-kt}D_{k+s,l+t}, \qquad k,l,s,t \in \mathbb{Z}_d,
\end{equation*}
from which we have
\begin{equation*}
	{\rm tr}D_{k,l}D_{s,t}^{\dagger}=d\delta_{k,s}\delta_{l,t}, \qquad k,l,s,t \in \mathbb{Z}_d. \label{Orth}
\end{equation*}
Therefore $\{D_{k,l}/\sqrt{d}:k,l\in \mathbb{Z}_d\}$ constitutes an orthonormal basis for the Hilbert space $L(\mathbb{C}^d)$ consisting of linear operators on the space $\mathbb{C}^d$, with the inner product defined as $\langle A|B \rangle={\rm tr}(A^{\dagger}B).$ The Heisenberg-Weyl operators play a crucial role in the stabilizer formalism of quantum computation, as shall be illustrated below \cite{Feng20241}. In this work, we shall propose a quantifier of complexity based on the interplay between the quantum state and the Heisenberg-Weyl operators.

The discrete Heisenberg-Weyl group 
\begin{eqnarray*}
	{\cal P}_d=\{\tau^j D_{k,l}: j\in \mathbb{Z}_{2d},k,l \in \mathbb{Z}_d\}
\end{eqnarray*}
 generated by the Heisenberg-Weyl operators  is a subgroup of the full unitary group ${\cal U}_d$ consisting of unitary operators on $\mathbb{C}^d$. Its normalizer
\begin{eqnarray*}
	{\cal C}_d=\{U_c\in {\cal U}_d: U_c{\cal P}_d U_c^\dagger = {\cal P}_d\} \label{Cli}
\end{eqnarray*}
in ${\cal U}_d$ is called the Clifford group consisting of the Clifford operators. 

The simultaneous eigenstate with eigenvalue $1$ of a maximal Abelian subgroup of the Pauli group ${\cal P}_d$ ($c{\bf 1}$ for $c\neq 1$ are not in the subgroup) is defined as the pure stabilizer state. Mixed stabilizer states are non-trivial probabilistic mixtures of pure stabilizer states. All other states are called magic states (non-stabilizer states). In the resource theory of magic, stabilizer states are usually recognized as free states and magic states are treated as a kind of resource in fault-tolerant quantum computation \cite{Niel2010,Gott1997,Gott1998}.

\subsection{Characteristic functions}
For a $d$-dimensional quantum system $\mathbb{C}^d$ with computational basis $\{|j\rangle : j\in \mathbb{Z}_d\}$, the characteristic function (Fourier transform) of an operator (e.g., state or observables) $A$ on $\mathbb{C}^d$ is defined as \cite{Dai2022,Gross2006}
\begin{equation}\label{chrrho}
	c_{k,l}(A)={\rm tr}(D_{k,l}A), \qquad k,l \in \mathbb{Z}_d.
\end{equation}
The characteristic function of an operator provides an equivalent characterization of the operator  since
\begin{equation}
	A=\frac1d\sum_{k,l=0}^{d-1} c^{\ast}_{k,l}(A)D_{k,l}.
\end{equation}
For a given quantum state $\rho$, considering its square root $\sqrt{\rho}$ and expanding  
\begin{equation}\label{sqrcha}
	\sqrt{\rho}=\frac1d\sum_{k,l=0}^{d-1} c^{\ast}_{k,l}(\sqrt{\rho})D_{k,l},
\end{equation}
then based on Eq. (\ref{sqrcha}), the quantum state $\rho$ can be expressed as 
\begin{eqnarray*}
	\rho &=& \sqrt{\rho} \cdot \sqrt{\rho}\\
	&=& \Big(\frac1d\sum_{k,l=0}^{d-1}c_{k,l}^{\ast}(\sqrt{\rho})D_{k,l}\Big)\cdot \Big(\frac1d\sum_{s,t=0}^{d-1}c_{s,t}^{\ast}(\sqrt{\rho})D_{s,t}\Big) \\
	&=& \frac{1}{d^2}\sum_{k,l,s,t=0}^{d-1}c_{k,l}^{\ast}(\sqrt{\rho})c_{s,t}^{\ast}(\sqrt{\rho})\tau^{ls-kt}D_{k+s,l+t}.
\end{eqnarray*}
Therefore the characteristic function of  $\sqrt{\rho}$ can fully specify the quantum state $\rho$. 
Since ${\rm tr}\rho=1$, we can deduce that the set of coefficients $\{c_{k,l}(\sqrt{\rho})\}$  satisfy the following constraint
\begin{equation}\label{const1}
	\sum_{k,l=0}^{d-1} |c_{k,l}(\sqrt{\rho})|^2=d.
\end{equation}

\subsection{SIC-POVMs and fiducial states}

Symmetric informationally complete measurements (SIC-POVMs) are generalized quantum measurements composed of $d^2$ equiangular complex lines in a $d$-dimensional Hilbert space. Recall that a SIC-POVM is defined by a set of POVM operators \cite{Rene2004}
$$\{E_\alpha: \alpha=1,2,\dots,d^2\}$$
which satisfies $E_\alpha \geq 0$, $\sum_{\alpha=1}^{d^2} E_\alpha ={\bf 1},$ and 

(1) Informationally completeness: $\{E_\alpha\}$ are rank-one operators which span the whole state space.

(2) Symmetry: ${\rm tr}E_{\alpha}={1}/{d}$ is a constant independent of $\alpha$, and ${\rm tr}(E_{\alpha}E_{\beta})={1}/\big(d^2(d+1)\big).$ 

Due to its informational completeness and high symmetry, SIC-POVMs have attracted lots of attention and have proven to be useful in both theoretical study of quantum foundations and practical applications \cite{SIC2006,SIC2010,Zhu2010,SIC2011,SIC2015,SIC2018,SIC2019,SIC2020}.

A pure state $|f\rangle\in \mathbb{C}^d$ is called a SIC-POVM fiducial state if its orbit under the discrete Heisenberg-Weyl group leads to a SIC-POVM in the sense that 
$$\left\{E_{k,l}=\frac1d D_{k,l}|f\rangle\langle f| D_{k,l}^{\dagger}: k, l\in \mathbb{Z}_d \right\}$$
is a SIC-POVM \cite{Zaun2011,Rene2004}. Despite their profound importance, the existence of SIC-POVM fiducial states  for all dimensions remains an outstanding open problem (Zauner’s conjecture) \cite{Zaun2011,Rene2004}.

\section{Quantify complexity of quantum states} \label{s3}

In this section, we introduce a quantifier of complexity of quantum states in terms of Jordan product and Lie product between the square root of the states and the Heisenberg-Weyl operators, and study its basic properties.

\subsection{Jordan product and Lie product}
For any quantum state $\rho$, we consider the following (symmetric) Jordan product between $\sqrt{\rho}$ and the Heisenberg-Weyl displacement operators $D_{k,l}$
$$\{D_{k,l},\sqrt{\rho}\}=D_{k,l}\sqrt{\rho}+\sqrt{\rho}D_{k,l},$$
and the (skew-symmetric) Lie product 
$$[D_{k,l},\sqrt{\rho}]=D_{k,l}\sqrt{\rho}-\sqrt{\rho}D_{k,l}.$$
Adopting the Hilbert-Schmidt norm of operators in the system Hilbert space 
$$\lVert X \rVert^2={\rm tr}(XX^{\dagger}),$$
we introduce the following two quantities
\begin{eqnarray}
	J(\rho,D_{k,l})&=&\frac12\lVert \{D_{k,l},\sqrt{\rho}\} \rVert^2,\\
	I(\rho,D_{k,l})&=&\frac12\lVert [D_{k,l},\sqrt{\rho}] \rVert^2.
\end{eqnarray}

Noticing that $I(\rho,D_{k,l})$ can be explicitly expressed as 
\begin{equation*}
	I(\rho,D_{k,l})=\frac12 {\rm tr}[\sqrt{\rho},D_{k,l}] [\sqrt{\rho},D_{k,l}]^{\dagger},
\end{equation*}
which is actually the generalized Wigner-Yanase skew information \cite{WignerYanase}. It may be regarded as a form of generalized quantum Fisher information \cite{hansen2008metric}, and has found widespread use in quantifying quantum coherence and deriving new uncertainty relations \cite{App1,App2,App3,App4,App5,App6}.

After straightforward algebraic manipulations, we arrive at
\begin{eqnarray}
\label{JRHO} J(\rho,D_{k,l})&=&1+{\rm tr}\left(\sqrt{\rho}D_{k,l}\sqrt{\rho}D_{k,l}^{\dagger}\right),\\
   \label{IRHO} I(\rho,D_{k,l})&=&1-{\rm tr}\left(\sqrt{\rho}D_{k,l}\sqrt{\rho}D_{k,l}^{\dagger}\right),  
\end{eqnarray}
and thus immediately obtain the corresponding trade-off relation
\begin{equation*}
	J(\rho,D_{k,l})+I(\rho,D_{k,l})=2.
\end{equation*}

This result manifests the complementary relation between these two quantities. Since $I(\rho,D_{k,l})$ captures the quantum characteristics of $\rho$ relative to $D_{k,l},$ it is natural to regard  $J(\rho,D_{k,l})$ as embodying the classical characteristics in terms of the commutative Jordan product. These two quantities are the core ingredients in quantifying complexity within the stabilizer formalism, as elaborated in the subsequent subsection.

\subsection{A quantifier of complexity}

While plentiful research has been devoted to quantum features such as non-stabilizerness (magic resource) \cite{Howa2017,Ahma2018,Hein2019,Wanf2019,Liu2022,Feng2022,Leon2022,Labi2022,Dai2022,Haug2023,Haug20232,Li2025,Zhan2026}, as well as  coherence \cite{App2,App3,App4,App5,App6,Gour2008,Baumgratz2014,Yu20161,Winter2016,Streltsov2017,Luo2017}, quantification of quantum state complexity has received comparatively less attention. In the stabilizer formalism, it seems plausible that a natural quantifier of quantum state complexity should satisfy the following desirable criteria: (1) it is neither convex nor concave; (2) it attains its minimum for stabilizer states and its maximum for SIC-POVM fiducial states; (3) it is invariant under Clifford operations.

For an arbitrary quantum state $\rho$, we propose an information-theoretic quantifier of its complexity in the stabilizer formalism as 
\begin{equation}\label{key}
	\mathcal{C}(\rho)=\sum_{k,l=0}^{d-1}I(\rho,D_{k,l})J(\rho,D_{k,l}).
\end{equation} 

After direct computation, it admits a succinct expression in terms of the characteristic functions of $\sqrt{\rho}$ as follows.
\vskip 0.2cm

{\bf Proposition 1.}  For $\mathcal{C}(\rho)$ defined by Eq. (\ref{key}), it holds that  
\begin{equation}\label{Key2} 
	\mathcal{C}(\rho)=d^2-\sum_{k,l=0}^{d-1}|c_{k,l}(\sqrt{\rho})|^4.
\end{equation}

To derive the above result, substituting the expression of $\sqrt{\rho}$ given by Eq. (\ref{sqrcha}) into Eqs. (\ref{JRHO}) and (\ref{IRHO}), we get 
\begin{eqnarray*}
	J(\rho,D_{k,l})&=&1+\frac1d\sum_{m,n=0}^{d-1}|c_{m,n}(\sqrt{\rho})|^2\omega^{lm-kn},\\
    	I(\rho,D_{k,l})&=&1-\frac1d\sum_{m,n=0}^{d-1}|c_{m,n}(\sqrt{\rho})|^2\omega^{lm-kn},
\end{eqnarray*}
therefore the complexity  defined by Eq. (\ref{key}) can be expressed as
 \begin{widetext}
 \begin{eqnarray*}
	\mathcal{C}(\rho)&=&\sum_{k,l=0}^{d-1}\Big(1-\frac1d\sum_{m_1,n_1=0}^{d-1}|c_{m_1,n_1}(\sqrt{\rho})|^2\omega^{lm_1-kn_1}\Big)\Big(1+\frac1d\sum_{m_2,n_2=0}^{d-1}|c_{m_2,n_2}(\sqrt{\rho})|^2\omega^{lm_2-kn_2}\Big)\\
	&=&\sum_{k,l=0}^{d-1}\Big(1-\frac{1}{d^2}\sum_{m_1,n_1,m_2,n_2=0}^{d-1}|c_{m_1,n_1}(\sqrt{\rho})|^2|c_{m_2,n_2}(\sqrt{\rho})|^2\omega^{l(m_1+m_2)-k(n_1+n_2)}\Big)\\
	&=&d^2-\frac{1}{d^2}\Big(\sum_{m_1,n_1,m_2,n_2=0}^{d-1}|c_{m_1,n_1}(\sqrt{\rho})|^2|c_{m_2,n_2}(\sqrt{\rho})|^2\Big)\Big(\sum_{l=0}^{d-1} \omega^{l(m_1+m_2)}\Big)\Big(\sum_{k=0}^{d-1}\omega^{-k(n_1+n_2)}\Big)\\
	&=&d^2-\frac{1}{d^2}\Big(\sum_{m_1,n_1,m_2,n_2=0}^{d-1}|c_{m_1,n_1}(\sqrt{\rho})|^2|c_{m_2,n_2}(\sqrt{\rho})|^2\Big)\Big(d\delta_{m_1+m_2,0}\Big)\Big(d\delta_{n_1+n_2,0}\Big)\\
	&=&d^2-\sum_{m,n=0}^{d-1}|c_{m,n}(\sqrt{\rho})|^4.
\end{eqnarray*}
\end{widetext}

The proposed complexity quantifier $\mathcal{C}(\rho)$  possesses several desirable properties, as listed below.

\vskip 0.2cm

{\bf Proposition 2}. For $\mathcal{C}(\rho)$ defined by Eq. (\ref{key}), it holds that 
		
	(a) Clifford invariance: $\mathcal{C}(\rho)=\mathcal{C}(U_c\rho U_c^{\dagger})$ for any Clifford unitary operator $U_c \in {\cal C}_d.$
	
    (b) Among pure quantum states,
    $$d^2-d\leq \mathcal{C}(|\phi\rangle\langle \phi|)\leq d^2-\frac{2d}{d+1},$$
    where the minimum is obtained if and only if the quantum state is a pure stabilizer state, and the maximum is achieved if and only if the quantum state is the SIC-POVM fiducial state (assuming their existence).

    (c) Among all (pure or mixed) quantum states, 
    $$0\leq \mathcal{C}(\rho)\leq d^2-\frac{2d}{d+1},$$
    where the minimum is obtained if and only if the quantum state is the maximally mixed state, while the maximum is achieved if and only if the quantum state is the SIC-POVM fiducial state (assuming their existence).

(d)  $\mathcal{C}(\rho)$ is not concave for all dimensions $d$, and is not convex for $d>2$ (The convexity in the case $d=2$ remains open).


\vskip 0.2cm

We proceed to prove the above results. 

For property (a), it is obvious since the Clifford operator permutes the Heisenberg-Weyl operators via conjugation.

For property (b), considering pure quantum states $\rho=|\phi\rangle \langle \phi|$, following Eq. (\ref{relati}), the proposed complexity quantifier given by Eq. (\ref{Key2}) can be expressed as
\begin{eqnarray*}
    \mathcal{C}(|\phi\rangle \langle \phi|)
    &=&d^2-\sum_{k,l}|c_{k,l}(|\phi\rangle \langle \phi|)|^4.
\end{eqnarray*}
In Ref. \cite{Feng2022}, it was shown that the upper bound of $M_4(|\phi\rangle \langle \phi|):=\big(\sum_{k,l}|c_{k,l}(|\phi\rangle \langle \phi|)|^4\big)^{1/4}$ is achieved if and only if $|\phi\rangle$ is a stabilizer state, while the lower bound is achieved if and only if $|\phi\rangle$ is a fiducial state (if it exists). Therefore the extreme values of $\mathcal{C}(|\phi\rangle \langle \phi|)$ can be directly calculated.

For property (c), among all quantum states, it is obvious that $\mathcal{C}(\rho)\geq 0$. The minimum is obtained if and only if $[D_{k,l},\sqrt{\rho}]=0$ for all $k,l=1,\cdots, d$. Noticing the fact that 
$$\left\{\frac{1}{\sqrt{d}}D_{k,l}:k,l\in \mathbb{Z}_d\right\}$$
constitutes an orthonormal basis for $L(\mathbb{C}^d)$, $\sqrt{\rho}$ can only be proportional to the identity operator, therefore $\rho$ is the maximally mixed state.

On the other hand, according to Eq. (\ref{const1}), we have
$$\sum_{(k,l)\neq (0,0)} |c_{k,l}(\sqrt{\rho})|^2=d-({\rm tr}\sqrt{\rho})^2.$$
Combining the above equation with the Cauchy-Schwarz inequality, we  get 
\begin{align*}
	& \sum_{(k,l)\neq (0,0)} |c_{k,l}(\sqrt{\rho})|^2 \\
&\leq \Big (\sum_{(k,l)\neq (0,0)} |c_{k,l}(\sqrt{\rho})|^4\Big )^{\frac12} \Big (\sum_{(k,l)\neq (0,0)} 1\Big )^{\frac12},
\end{align*}
which can be equivalently expressed as
\begin{eqnarray*}
	\sum_{(k,l)\neq (0,0)} |c_{k,l}(\sqrt{\rho})|^4 &\geq& \frac{\left(d-({\rm tr}\sqrt{\rho})^2\right)^2}{d^2-1},
\end{eqnarray*}
therefore
\begin{align*}
	\mathcal{C}(\rho)&=d^2-\sum_{m,n}|c_{m,n}(\sqrt{\rho})|^4\\
	&=d^2-({\rm tr}\sqrt{\rho})^4-\sum_{(m,n)\neq (0,0)}|c_{m,n}(\sqrt{\rho})|^4\\
	&\leq d^2-({\rm tr}\sqrt{\rho})^4-\frac{\left(d-({\rm tr}\sqrt{\rho})^2\right)^2}{d^2-1}\\
	&\leq d^2-1-\frac{(d-1)^2}{d^2-1}\\
	&= d^2-\frac{2d}{d+1},	
\end{align*}
where the maximum is achieved if and only the quantum state is pure and $|c_{m,n}(\sqrt{\rho})|$ equals for all $(m,n)\neq (0,0)$. Therefore the maximum is achieved if and only if the quantum state is the SIC-POVM fiducial state.

For property (d), $\mathcal{C}(\rho)$ is clearly not concave since for any dimension $d,$
$$\mathcal{C}\Big(\frac{\bf 1}{d}\Big)=0 <\sum_{k\in{\mathbb Z}_d} \frac1d \mathcal{C}(|k\rangle\langle k|)=d^2-d,$$
where ${\bf 1}/{d}=\sum_{k\in{\mathbb Z}_d} |k\rangle\langle k|/d$ is the maximally mixed state and $\{|k\rangle, k\in{\mathbb Z}_d\}$ are pure stabilizer states.

Regarding the convexity of $\mathcal{C}(\rho)$, consider the family of mixed states
$$\rho_p=p|\psi\rangle\langle\psi|+(1-p)\frac{\bf 1}{d},\qquad p\in [0,1],$$
where $|\psi\rangle$ is an arbitrary pure state. Its square root is
$$\sqrt{\rho_p}=\Big(\sqrt{\frac1d+(1-\frac1d)p}-\sqrt{\frac1d-\frac{p}d}\Big)|\psi\rangle\langle\psi|+\sqrt{\frac1d-\frac{p}d}{\bf 1},$$
and the corresponding complexity is 
\begin{align*}
&\ \mathcal{C}(\rho_p)\\
&=d^2-\Big(\sqrt{\frac1d+(1-\frac1d)p}+(d-1)\sqrt{\frac1d-\frac{p}d}\Big)^4\\
&-\Big(\sqrt{\frac1d+(1-\frac1d)p}-\sqrt{\frac1d-\frac{p}d}\Big)^4\big(d^2-1-\mathcal{C}(|\psi\rangle\langle\psi|)\big),
\end{align*}
where $\mathcal{C}(|\psi\rangle\langle\psi|)$  denotes the complexity of the pure state $|\psi\rangle.$

For $|\psi\rangle$ being an arbitrary pure stabilizer state (so that $\mathcal{C}(|\psi\rangle\langle\psi|)=d^2-d$),  the second derivative at $p=0$ is positive, that is,
\begin{align*}
\frac{\partial^2\mathcal{C}(\rho_p)}{\partial p^2}\Big|_{p=0}=d^2(d-1)>0.
\end{align*}
By expanding near $p=1-,$ we obtain 
\begin{align*}
\mathcal{C}(\rho_p)&=d^2-d-4(d-1)(1-p)\\
&-\frac{4(d-1)(d-2)}{\sqrt{d}}(1-p)^{3/2} +O\big((1-p)^{2}\big),
\end{align*}
and the second derivative
\begin{align*}
\frac{\partial^2\mathcal{C}(\rho_p)}{\partial p^2}\Big|_{p=1-}=-\frac{3(d-1)(d-2)}{\sqrt{d(1-p)}} +h_d+O\big(\sqrt{1-p}\big)
\end{align*}
with $h_d=2(8-16d+9d^2-d^3)/d,$ diverges to $-\infty$ for $d>2.$ Since $\mathcal{C}$ is a smooth function of $p$ on $(0,1)$, its second derivative takes a positive value at $p=0$ and diverges to $-\infty$ as $p\to 1-$, therefore there must exist a region near $p=1-$ where the second derivative is negative, which implies a violation of convexity for $d>2.$ As a concrete example, take $d=3$ and $|\psi\rangle$ being a pure stabilizer state. Then $\rho_{0.95}=(\rho_{0.9}+\rho_1)/2=(\rho_{0.9}+|\psi\rangle\langle\psi|)/2,$ $\mathcal{C}(|\psi\rangle\langle\psi|)=6,$ and we have
$$\mathcal{C}(\rho_{0.95})\approx5.5609>\frac12\big(\mathcal{C}(\rho_{0.9})+\mathcal{C}(\rho_1)\big)\approx5.5528,$$
explicitly demonstrating that $\mathcal{C}(\rho)$ is not convex for $d=3$.

The above analysis fails for $d=2$ as the second derivative is finite and positive, consistent with convexity in this example. We randomly sampled one hundred thousand probabilistic mixtures of qubit states and found no counterexamples of convexity. Thus, numerical evidence indicates that $\mathcal{C}(\rho)$ is convex for $d=2.$ However, an analytic proof is still lacking.

\vskip 0.2cm

Several remarks are in order. On the one hand, the maximally mixed state emerges as the least complex state among all quantum states, a property that is highly intuitive with respect to the quantification of complexity. By contrast, stabilizer states—which act as classical analogs in the stabilizer formalism of quantum computation and can simultaneously be mathematically well-characterized—represent the least complex states among all pure quantum states. On the other hand, SIC-POVM fiducial states correspond to the most complex states. Mathematically, their existence for all dimensions remains an outstanding open problem (Zauner’s conjecture), which has garnered considerable attention in the field. Physically, these states may have important implication for stabilizer quantum computation. Here, we have offered a novel characterization of SIC-POVM fiducial states from a complexity-theoretic perspective.

\begin{table*}[!htbp]
\renewcommand\arraystretch{1.5}
\caption{\label{table1} Comparison of the quantifiers of magic $M_p(\rho) \ (p\geq 2)$  and complexity $\mathcal{C}(\rho)$ among pure quantum states}
\begin{ruledtabular}
\begin{tabular}{ccccc}
Quantifiers: & \multicolumn{2}{c}{Quantifier of magic $M_p(\rho) \ (p\geq 2)$} & \multicolumn{2}{c}{Quantifier of complexity $\mathcal{C}(\rho)$} \\ \hline
Extremal states:      & fiducial states (min)   & stabilizer states (max) & stabilizer states  (min) &  fiducial states (max)\\ \hline
Extremal values: & $(1+(d-1)(d+1)^{1-p/2})^{1/p}$   &$d^{1/p}$   & $d^2-d$     & $d^2-2d/(d+1)$ \\
\end{tabular}
\end{ruledtabular}
\end{table*}

\subsection{Relationship with magic (non-stabilizerness)}

In the stabilizer formalism, magic is the key resource which has attracted lots of attention recently \cite{Howa2017,Ahma2018,Hein2019,Wanf2019,Liu2022,Feng2022,Leon2022,Labi2022,Dai2022,Haug2023,Haug20232,Li2025,Zhan2026}. In this subsection, we establish a relationship between complexity and magic. 

Recall that in Refs. \cite{Dai2022,Feng2022},  the $L^p$-norm of the characteristic function 
\begin{equation}\label{magicqua}
	M_p(\rho)=\Big (\sum_{k,l}|c_{k,l}(\rho)|^p\Big )^{1/p}
\end{equation}
as a quantifier (witness) of magic was investigated. 
Comparing Eq. (\ref{magicqua}) with Eq. (\ref{Key2}), we can see the subtle differences and similarities between these two quantifiers. Explicitly, both these two quantities are based on the characteristic functions. Yet one is based on the characteristic function of the quantum state $c_{k,l}(\rho)$, while the other one is based on the characteristic function of the square root of the quantum state $c_{k,l}(\sqrt{\rho})$. 

Since $\sqrt{\rho}=\rho$ for pure states, the following complementarity relation 
\begin{equation}\label{relati}
    M_4^4(\rho)+\mathcal{C}(\rho)=d^2
\end{equation}
holds for any pure state $\rho$. Correspondingly, the extremal states among the pure quantum states are the stabilizer states and fiducial states, respectively, as listed in Table \ref{table1}.

In this context and inspired by Eq. (\ref{magicqua}), it is reasonable to study the following quantity
\begin{equation}\label{ma}
	{\cal M}_p(\rho)=\Big (\sum_{k,l}|c_{k,l}(\sqrt \rho)|^p\Big )^{1/p}
\end{equation}
as a quantifier of magic or (un)complexity.

\section{Illustrative examples} \label{s4}

In this section, we illustrate the quantifier of complexity for the simplest quantum system $\mathbb{C}^2$. 

Taking the given computational basis $\{\ket{0},\ket{1}\}$, the shift operator $X$ and the phase operator $Z$ can be explicitly expressed as
\begin{eqnarray*}
X=\sigma_x=\bigg (\begin{array}{cc}
                          0 & 1 \\
                          1 & 0 \\
                        \end{array}\bigg ),  \qquad
Z=\sigma_z=\bigg (\begin{array}{cc}
                          1 & 0 \\
                          0 & -1 \\
                        \end{array}\bigg ),
\end{eqnarray*}
which are the Pauli $X$ and Pauli $Z$ operators, respectively. Together with the identity operator ${\bf 1}$ and the Pauli $Y$ operator
\begin{eqnarray*}
\sigma_y=\bigg (\begin{array}{cc}
                          0 & -i \\
                          i & 0 \\
                        \end{array}\bigg ),
\end{eqnarray*}
they constitute the conventional Pauli matrices. The density operator can be uniquely expressed as
\begin{equation*}
	\rho=\frac{1}{2}({\bf 1}+{\boldsymbol{r}}\cdot{\bf \sigma})\label{rho},
\end{equation*}
where ${\bf 1}$ is the identity operator, ${\boldsymbol\sigma}=(\sigma_{x},\sigma_{y},\sigma_{z})$ are the Pauli matrices, and ${\boldsymbol{r}}=(r_{1},r_{2},r_{3})\in\mathbb{R}_3$ is the Bloch vector of $\rho$ which satisfies 
\begin{eqnarray*}
r=||{\boldsymbol{r}}||=\sqrt{r_1^2+r_2^2+r_3^2} \in[0,1].
\end{eqnarray*}
For pure quantum states, it holds that $r=1$, whereas for all mixed quantum states, $r<1.$ 

The four Heisenberg-Weyl operators are 
\begin{equation*}
    D_{0,0}={\bf 1},\  D_{0,1}=\sigma_z, \ D_{1,0}=\sigma_x,\  D_{1,1}=-i\sigma_y,
\end{equation*}
respectively. 
After substituting the displacement operators into Eq. (\ref {key}) and conducting direct calculations, the complexity quantifier of $\rho$ can be explicitly expressed as
\begin{equation}
 \mathcal{C}(\rho)=4-s^2-\frac{r_1^4+r_2^4+r_3^4}{s^2}  
\end{equation}
with $s=1+\sqrt{1-r^2}$.

We can explicitly analyze the extreme values and their corresponding extremal states.

(1) Among pure quantum states $|\phi\rangle\langle \phi|$, since $r=1$, $s=1$, we have
$$\mathcal{C}(|\phi\rangle\langle \phi|)=3-(r_1^4+r_2^4+r_3^4).$$
Mathematically, we have
\begin{eqnarray*}
\frac{1}{3}\left(r_{1}^{2}+r_{2}^{2}+r_{3}^{2}\right)^2\leq r_{1}^{4}+r_{2}^{4}+r_{3}^{4} \leq \left(r_{1}^{2}+r_{2}^{2}+r_{3}^{2}\right)^2,
\end{eqnarray*}
therefore among all pure qubit states,
\begin{eqnarray*}
2 \leq \mathcal{C}(|\phi\rangle\langle \phi|)\leq \frac{8}{3}.
\end{eqnarray*}

The pure states which achieve the minimum of $\mathcal{C}(|\phi\rangle\langle \phi|)$ correspond to the following Bloch vectors
\begin{eqnarray*}
{\boldsymbol r}_S\in\big\{(0,0,\pm 1),(0,\pm 1,0),(\pm 1,0,0)\big\}.
\end{eqnarray*}
These are actually the six eigenvectors of the Pauli matrices $\{\ket{0_\alpha},\ket{1_\alpha},\alpha=x,y,z\}$.
While the Bloch vectors of pure states which achieve the maximal value of  $\mathcal{C}(|\phi\rangle\langle \phi|)$ are
\begin{eqnarray*}
{\boldsymbol r}_T\in \Big \{\Big(\pm\frac{1}{\sqrt{3}},\pm\frac{1}{\sqrt{3}},\pm\frac{1}{\sqrt{3}}\Big)\Big \},
\end{eqnarray*}
the corresponding states are the $T$-type magic states
\begin{eqnarray*}
\ket{T}\!\bra{T}=\frac{1}{2}\Big({\bf1}+\frac{1}{\sqrt{3}}(\pm\sigma_x\pm\sigma_y\pm\sigma_z)\Big),
\end{eqnarray*}
which are exactly the SIC-POVM fiducial states in the qubit system.

(2) For general qubit states, we first analyze the situation when $r$ is fixed. Correspondingly, $s$ is fixed. Similarly, we have
$$s^2+\frac{r^4}{3s^2}\leq s^2+\frac{r_1^4+r_2^4+r_3^4}{s^2} \leq s^2+\frac{r^4}{s^2}.$$
Therefore, 
\begin{equation}\label{genr}
 4-s^2-\frac{r^4}{ s^2}\leq \mathcal{C}(\rho)\leq 4-s^2-\frac{r^4}{3s^2}.  
\end{equation}

The upper bound in Eq. (\ref{genr}) increases as $r$ increases. Therefore it achieves its maximum for pure quantum states which satisfies $$r_1^2=r_2^2=r_3^2=\frac13.$$ These are exactly the $T$-type magic states. Meanwhile, for the maximally mixed state,
$$\mathcal{C}\Big(\frac{\bf 1}{2}\Big)=0,$$ 
which corresponds to the minimal value of $\mathcal{C}(\rho)$.

\section{Summary}\label{s5}

In this work, we have analyzed the complexity within the stabilizer formalism and defined it as the sum, over all phase-space points, of the product of the squared norms of the anticommutator and commutator between $\sqrt{\rho}$ and the Heisenberg-Weyl operators. These two quantities capture the quantum and classical features in the characterization of complexity: one can be regarded as a generalized quantum Fisher information, while the other is complementary to it.

Remarkably, the proposed quantifier admits a compact and physically insightful expression solely in terms of the fourth moments of the characteristic function of  $\sqrt{\rho}.$ We have revealed that this quantifier possesses several desirable properties. Furthermore, the extremal states with respect to this complexity measure are precisely the canonical families of states in the stabilizer formalism, including pure stabilizer states and SIC-POVM fiducial states (assuming their existence). This, in turn, offers a novel characterization of these extremal states from the perspective of statistical complexity.

Our findings establish a direct and natural connection between complexity and the geometric symmetry of SIC-POVMs. Beyond providing a computationally tractable quantifier of complexity within the stabilizer formalism, this work deepens our understanding of quantum complexity by linking it to fundamental structures of quantum information science. It would also be worthwhile to introduce a similar quantity in the context of continuous-variable quantum systems, one that differs from existing complexity measures grounded in phase-space representations and probability distributions. We leave this for future work.

\section*{Acknowledgements}

This work was supported by the Fundamental Research Funds for the Central Universities, Grant No. FRF-TP-19-012A3, the Youth Promotion Association of CAS Grant No. 2023004, the National Natural Science Foundation of China, Grant Nos. 12401609 and 12341103.

\end{document}